\documentclass[prb,aps]{revtex4}
\usepackage{graphicx}
\newcommand{\normwidth}{0.7\columnwidth}

\begin{document}

\title{Pumping in an interacting quantum wire}

\author{R. Citro}
\affiliation{Dipartimento di Fisica ``E. R. Caianiello'' and
Unit{\`a} I.N.F.M. di Salerno\\ Universit{\`a} di Salerno, Via S.
Allende, 84081 Baronissi (Sa), Italy}
\author{ N. Andrei}
\affiliation{Center for Materials Theory, Serin Physics Laboratory, Rutgers University\\
          Piscataway, New Jersey 08854-8019, USA}
\author{Q. Niu}
\affiliation{Department of Physics, The University of Texas\\
          Austin, Texas 78712, USA}
\begin{abstract}
We study charge and spin pumping in an interacting one-dimensional
wire. We show that a spatially periodic potential modulated in space and time
 acts as a quantum pump inducing a
dc--current component at zero bias. The current generated by the
pump is strongly affected by the interactions. It has a power law
dependence on the frequency or temperature with the exponent
determined by the interaction in the wire, while the coupling
to the pump affects the amplitudes only. We also show that pure
spin-pumping can be achieved, without the presence of a magnetic
field.
\end{abstract}


\maketitle

\section{Introduction}\label{sec:introduction}

 An adiabatic quantum pump is a device that generates a dc--current (at
zero bias) by a periodic slow variation of some system
characteristic, the variation being slow enough so that the system
remains close to its ground state throughout the pumping cycle.
The physics of pumping has attracted considerable interest in the
last two decades: In his original work Thouless\cite{thouless_or}
studied the integrated particle current on a finite torus produced
by a slow variation of the potential and showed that the integral
of the current over a period can vary continuously, but must have
an integer value in a clean infinite periodic system with full
bands.  The robustness of the quantization in the latter system
with respect to the influence of disorder, many-body interactions
and system size was shown in Refs.[\onlinecite{niu,niudis}], and
spectacular precision of quantization of the pumped current has
also been achieved in experiment\cite{shilton_exp}.  Since then,
interest in this phenomenon has shifted to
theoretical\cite{brouwer,aleiner1,polianski,zhou} and
experimental\cite{switches,altsh} investigations of adiabatic
pumping through open quantum dots where the realization of the
periodic time-dependent potential can be achieved by modulating
gate voltages applied to the structure.  In this regime, the
pumped current is generally not
quantized\cite{levinson,aharony_qd}, and interesting questions are
raised on the nature of dissipation associated with the
pumping\cite{aleiner2,buttiker_noise,polianski_noise,avron_qp}.
Recently, theoretical studies of quantum pumping have extended to
systems with exotic leads, such as superconductor
wires\cite{wang_sup_qp,wang_sup_op_qp} and Luttinger liquid
quantum wires\cite{chamon,chamon_long}.  A single-wall carbon
nanotube represents an ideal realization of such an interacting
quantum wire, and parametric pumping can be achieved by applying
gate voltages on the sides\cite{wei_carbon} or surface acoustic
wave propagating along the wire \cite{acwaves1}.

\begin{figure}
\centerline{\includegraphics[angle=-0,width=\normwidth]{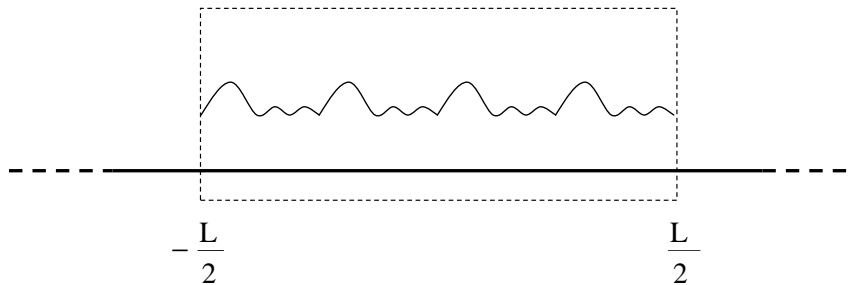}}
\vspace*{-8.cm}
\caption{Quantum wire in presence of a periodic potential extending
from $-L/2$ to $L/2$ and oscillating with frequency $\omega_0$.}
\label{fig:qw}
\end{figure}

In this paper, we report our results on quantum pumping through an
interacting one-dimensional wire in the adiabatic regime.  The pump we
propose consists of a spatially periodic potential $V$ extending from
$-L/2$ to $+L/2$ and oscillating wave-like  with frequency
$\omega_0$ and momentum $q_0$, acting on an interacting clean quantum
wire of infinite length, see Fig.\ref{fig:qw}.  We shall show that
d.c. spin and charge currents are induced.

The low energy properties of the quantum wire are described by a
Luttinger liquid, the fixed point hamiltonian of the wire, and we
carry out the pumping at low temperatures and small $\omega_0$,
staying this way in the neighborhood of the fixed point. In this
regime, the charge is not quantized as expected, and the results
reflect the intrinsic properties of the Luttinger liquid.  An
anomalous response will be observed since the external periodic potential
couples to electrons while the quasiparticles of the interacting
systems are Luttinger-like bosons. We will also address the issue of a
pure spin pumping through an interacting quantum wire.

The paper is organized as follows. In Sec.\ref{sec:hamiltonian} we
introduce our physical setup and the Hamiltonian that describes
the pump in a 1-dimensional wire, making use of the Luttinger
liquid description. In Sec.\ref{sec:noneq} we introduce the
non-equilibrium Keldysh formalism appropriate to calculate the
charge and spin current in the wire. After that we discuss the
results for the current at zero and finite temperature. Finally we
draw the conclusions in Sec.\ref{sec:conclusions} discussing
further perspectives of our analysis and the implications for the
experimental realization of a device.

\section{The Hamiltonian of a pumped 1-d wire}\label{sec:hamiltonian}

There are several experimental realizations of  1-dimensional
systems, among which  are nanotubes, quantum wires and
organic conductors, such as Bechgaard salts. These systems are
described by interacting 1-dimensional hamiltonians, generally of the form:

\begin{eqnarray}\label{1dham}
&&H_T= H_0 + H_{el,el}\\
&&H_0\;\; = \;\sum \epsilon_k c^{\dagger}_{k \sigma} c_{k \sigma} \\
&&H_{el,el}= \sum_{k \in BZ} U^{\sigma_1\sigma_2\sigma_3\sigma_4}_{(k_1,k_2,k_3,k_4)} c^{\dagger}_{k_1\sigma_1}c^{\dagger}_{k_2 \sigma_2} c_{k_3\sigma_3} c_{k_4\sigma_4}
\end{eqnarray}
\noindent where $c^\dagger_{k\sigma}$ is an electron creation operator
with momentum $k$ and spin component $\sigma$, $c_{j\sigma}^\dagger$ (its
Fourier transform) creates the electron at lattice site $x_j = ja$,
$U$ is an arbitrary electron-electron interaction.

If we wish to study the  low energy physics of such a model, as we
wish to do in the case of {\em adiabatic} pumping,  it suffices to
consider it close to its fixed point - typically the Luttinger
liquid - to which it flows under the action of the renormalization
group (RG)
 \cite{haldane}. The low energy dynamics takes place close to
the Fermi points $\pm k_F$ and is expressed in terms of the
fermionic low energy fields $\psi_{\alpha \sigma}(x)$ describing the
right moving modes ($\alpha =R$) with spin $\sigma$ around $k_F$ and
the left moving modes ($\alpha =L$) describing the physics
around $- k_F$. The  Luttinger hamiltonian is,
\begin{equation}\label{hll}
H_{LL}= -iv_F \int dx (\Psi^\dagger_{R\sigma}(x)  \partial_x \Psi_{R\sigma}(x)
-\Psi^\dagger_{L\sigma}(x)  \partial_x \Psi_{L\sigma}(x))+ g
\int dx \rho(x)^2,
\end{equation}
\noindent where $v_F$ is the Fermi velocity, $g$ measures the strength
of interactions ($g >0 $ for repulsive interactions),
$\rho=\rho_R+\rho_L$ is the sum of the left and right moving electron
densities $\rho_{\alpha, \sigma}=\Psi^\dagger_{\alpha
\sigma}\Psi_{\alpha \sigma}$, with
$\rho_\alpha=\sum_{\sigma}\rho_{\alpha, \sigma}$. Note that the number
of left and right movers is conserved by the Luttinger hamiltonian.

Consider now the  wire in the presence of an external periodic
potential. We add then to Eqn.(\ref{1dham}) the term,
\begin{eqnarray}
 H^{latt}= \sum_j V_{\sigma,\sigma'}(x_j) c^{\dagger}_{j,\sigma}
c_{j,\sigma'} \label{potential}
\end{eqnarray}
where $V_{\sigma,\sigma'}(x_j)$ is a periodic external potential
of $x_j$ (with period $l$) acting on a section of length $L$ of
the wire. A possible way to realize the periodic potential is to embed a
section  $L$ of the long qunatum wire in a
semiconductor heterostructure with a meander line on top (or
bottom) of the sandwich generating a spatial periodic electric
field oscillating in time with a fixed frequency $\omega_0$ at the
interface.  The interfacial electric field would be such that the
effective potential experienced by the Luttinger bosonic-like
quasiparticles will result in a sinusoidal potential modulated in
space and time. (In such system magnetized contacts could be used
to preferentially inject and detect specific spin
orientation\cite{das_spinemitter,mireles}.)

Also the periodic potential will flow under the action
of the renormalization group, and in the low energy limit it will
be represented by a sum over umklapp operators $H^U_{n,m,n_s}$
transferring $n$ electrons and $n_s$ units of spin from right to
left Fermi points (and vice versa), while absorbing from the
lattice $m$ units of lattice momentum $G = 2 \pi /l $
\cite{rosch}.
The umklapp operators to which (\ref{potential}) flows under RG
describe high-energy processes which are irrelevant (in the RG
sense) at low energies when we consider systems close to a
Luttinger fixed point. However, we shall examine the system at
small but finite energy scales at which the RG flow stops and the
umklapp terms make the main contribution to pumping.

Leading umklapp terms are of the form:

\begin{eqnarray}\label{hu}
&& H^U_{0,m,0} \approx g^U_{0,m,0} \int dx \lbrack e^{i \Delta
k_{0,m}x}(\rho_R+\rho_L)^2+h.c. \rbrack \nonumber \\
 && H^U_{1,m,0} \approx
g^U_{1,m,0} \sum_\sigma \int dx \lbrack e^{i \Delta k_{1,m}x}\Psi^\dagger_{R
\sigma}(x)\Psi_{L\sigma}(x)\rho_{-\sigma}+h.c. \rbrack \nonumber
\\
&& H^U_{1,m,1} \approx
g^U_{1,m,1}  \int dx \lbrack e^{i \Delta k_{1,m}x}\Psi^\dagger_{R
\downarrow}(x)\Psi_{L\uparrow}(x)\rho +h.c. \rbrack \nonumber
\\
&& H^U_{2,m,0} \approx g^U_{2,m,0} \int dx \lbrack e^{i \Delta
k_{2,m}x}\Psi^\dagger_{R\uparrow}(x)\Psi^\dagger_{R\downarrow}(x)
\Psi_{L\downarrow}(x)\Psi_{L\uparrow}(x) +h.c. \rbrack
\end{eqnarray}

\noindent with $\Delta k_{n,m}=n2k_F-m G$ being the momentum
transfer associated with the process $n,m$. Note that a
commensurability between the electron density and the imposed
periodicity implies $\Delta k_{n,m}=0$ for some $n,m$. At
commensurate filling some umklapp operator may become relevant.
This is the case with $H^U_{2,1,0}$ at half filling for any value
of the coupling $g^U_{2,1,0}$. This would also be the case with
other commensurate fillings, but with a finite critical value of
the coupling. When any of the umklapp operators is relevant the
low energy behavior is no longer given by the Luttinger liquid. We
shall assume in what follows that we are away from half filling,
and when considering other commensurate filling, that the coupling
is below its critical value.

Also boundary terms may be generated under the RG process. The periodic
potential acts on a section of the wire and we assumed sharp edges at
$\pm L/2$, hence terms  of the form,
\begin{eqnarray}
H^{boundary} = V_0 [ \psi^{\dagger}_{R \sigma}(L/2) \psi_{L \sigma}(L/2) +
\psi^{\dagger}_{R \sigma}(-L/2) \psi_{L \sigma}(-L/2) + h.c. ]
\end{eqnarray}
will appear. Such terms were shown by Kane and Fisher to be
relevant in the low energy limit\cite{kane_fisher}.

\bigskip

We now allow the external periodic potential to oscillate with
frequency $\omega_0$ and propagate with some momenta $\{ q \}, \; q \approx
q_0 + \delta q, \;  {\rm with} \; \delta q \ll q_0 $,

\begin{eqnarray}
V(x) \to V(t, x)= \sum_q A_q  \cos(\omega_0 t - q x)V(x).
\end{eqnarray}

Again, close to
the Luttinger fixed point, the potential renormalizes to a sum of
umklapp terms with time (and phase) dependent coupling constants:
\begin{eqnarray}
\label{timepot} g^U_{n,m}(t) = g^U_{n,m} e^{i (\omega_0 t
-\varphi_{n,m})}.
\end{eqnarray}
 The momenta $\{ q \}$ in the driving potential break the mirror
symmetry of the oscillating potential and are reflected in the
effective low energy hamiltonian by the umklapp phases
$\varphi_{n,m}$. For very weak periodic potential one expects
$\varphi_{n,m}\approx n q_0/\omega_0 $. When mirror symmetry is
present $\varphi_{n,m}=0$ (and we shall see that no current is
induced). Together with the periodic potential also the boundary
terms will oscillate and we have for the leading term,
$H^{boundary}(t) = V_0 [e^{i \omega_0 t } \psi^{\dagger}_{R
\sigma}(L/2) \psi_{L \sigma}(L/2) + e^{i (\omega_0 t - \varphi)}
\psi^{\dagger}_{R \sigma}(-L/2) \psi_{L \sigma}(-L/2) +h.c ]$,
where $\varphi$ is the temporal phase shift between the two edges.

The  low energy effective Hamiltonian,
\begin{eqnarray}
&& H_{eff}(t)=H_{LL}+ H^{pump}(t) \\
&& H^{pump}(t)= H^{bulk}(t)  + H^{boundary}(t) \\
&& H^{bulk}(t)= \sum_{m,n,n_s} H^U_{n,m,n_s}(t)
\end{eqnarray}
\noindent describes the time evolution of the system close to the
fixed point, and is valid therefore (over a cycle) when all energy
scales such as $\omega_0, T $ are small.  We shall show that the
oscillating potential acts as a quantum pump, inducing spin and
charge d.c. currents. We shall find that both the bulk term
$\sum_{m,n,n_s} H^U_{n,m,n_s}(t)$ and the boundary term
$H^{boundary}(t)$ induce charge and spin currents.  The bulk
contribution  dominates in the large pump limit, i.e. for $L \to
\infty$, holding $\omega_0$ fixed but small. In the other limit,
$\omega_0 \to 0$ holding $L$ large but fixed, the boundary
contribution dominates.

\bigskip

We wish to study the effect of the oscillating terms on the
current operators,
\begin{eqnarray}
I_c(x) &=&
\sum_\sigma \left (\psi^\dagger_{R\sigma}(x)\psi_{R\sigma}(x)
-\psi^\dagger_{L\sigma}(x)\psi_{L\sigma}(x)\right )\\
I_s(x) &=& \sum_{\sigma,\sigma'}\left (\psi^\dagger_{R \sigma} \tau^z_{\sigma \sigma'}(x)
\psi_{R\sigma'}(x)-\psi^\dagger_{L \sigma} \tau^z_{\sigma \sigma'}(x)
\psi_{L\sigma'}(x) \right )
\end{eqnarray}
To do so it  is convenient to rewrite the problem in terms of bosonic fields
$\phi_{\sigma}, \Pi_{\sigma}$:
Defining the chiral components $\phi_{R \sigma},\phi_{L \sigma}= 1/2 \left(
\phi_{\sigma}\pm \int^x \Pi_{\sigma}(x') d x'\right)$, the
fermionic fields are given by,
\begin{eqnarray}\label{field}
&&\psi_{R,\sigma}(x,t)=
\frac{1}{\sqrt{2 \pi a }} e^{i \phi_{R \sigma}}  \nonumber \\
&&\psi_{L,\sigma}(x,t)= \frac{1}{\sqrt{2 \pi a}} e^{-i \phi_{L \sigma}}.
\end{eqnarray}
\noindent where $a$ is a spatial cut off (essentially the {\it
electron} lattice spacing, to be distinguished from $l$). Rewriting
the interacting Hamiltonian Eqn.(\ref{hll}) by means of the bosonic
fields, it can be brought into a quadratic form by a Bogliubov
rotation\cite{solyom,emery}. It is convenient to introduce the
combinations $\phi_c=(\phi_\uparrow +\phi_\downarrow)/\sqrt{2}$ and
$\phi_s=(\phi_\uparrow -\phi_\downarrow)/\sqrt{2}$, the spin and
charge bosonic fields, in terms of which
\begin{equation}\label{ham}
H_{LL}=\frac{1}{2\pi} \sum_{\nu=c,s} v_\nu \int dx
\left(K_{\nu}\Pi_{\nu}^2 +\frac{1}{K_\nu}(\partial_x
\phi_\nu)^2 \right),
\end{equation}
\noindent where the momenta $\Pi_\nu$ are conjugate to $\phi_\nu$,
$v_{c,s}$ are the charge and spin velocities and $K_\nu$ are the
Luttinger parameters, $v_c/K_c=v_F+g/\pi$ and $v_s/K_s=v_F-g/\pi$.
The bosonic version of the umklapp terms  is,

\begin{equation}\label{huu}
H^U_{n,m, n_s}(t)= \frac{g^U_{n,m, n_s}}{(2\pi
a)^n} \int dx \{e^{i(\omega_0 t-\varphi_{n,m})} e^{i\Delta k_{n,m}x}
e^{i\sqrt{2}(n\phi_c +n_s \phi_s)}+h.c. \},
\end{equation}
while the local boundary term is,
\bigskip
\begin{equation}
H^{boundary}(t)=\frac{V}{(2\pi a)}[e^{i\omega _{0}t}e^{i\sqrt{2}(\phi _{c}(\frac{L}{%
2},t)+\phi _{s}(\frac{L}{2},t))}+e^{i(\omega
_{0}t-\varphi)}e^{i\sqrt{2}(\phi _{c}(-\frac{L}{2},t)+\phi
_{s}(-\frac{L}{2},t))}+h.c.\}.
\end{equation}

In terms of bosonic fields the  charge current and spin current are given by:
\begin{eqnarray}\label{irho}
I_c (x,t)&=&\frac{e\sqrt{2}}{\pi} \partial_t \phi_c (x,t) \nonumber
\\
I_s (x,t)&=&\frac{\hbar\sqrt{2}}{\pi} \partial_t \phi_s(x,t).
\end{eqnarray}
\noindent where $e$ denotes the electric charge.  In the
following, we shall consider the oscillating lattice as a
perturbation around the Luttinger liquid fixed point and compute
the current perturbatively. This is a controlled expansion in the
low energy limit as noted before. As we will show, the boundary
term, though relevant with respect to the Luttinger liquid, as
$\omega_0 \to 0$ will lead to a subdominant contribution in the
large pump limit.

\section{Non-equilibrium transport formalism}\label{sec:noneq}

In the system described above we consider an external source
pumping energy into it, therefore the general formalism of this
non-equilibrium situation is given by the Keldysh
technique\cite{keldysh}. Our purpose is to calculate the charge
and spin currents generated by the pumping. They are  given by:

\begin{equation}\label{meanc}
\langle I_{c,s}(x,t)\rangle=\langle T_C \{I_{c,s} (x,t)
e^{-i \oint dt_1 H^{pump}(t_1)}\}\rangle,
\end{equation}

\noindent where $T_C$ is the time ordering operator along the
Keldysh contour. Expressing $T_C$ in terms of the ordering/anti-ordering operator $T_K$
along the upper/lower Keldysh branches, we
adopt the convention\cite{crepieux} that the indices $\eta,\eta_{1,2}=\pm$
identify the upper/lower branch of the Keldysh contour.

We shall begin by studying the bulk contribution of the pump. We then
expand in the irrelevant umklapp operators around the Luttinger Liquid
fixed point.  Expanding the exponential to first order we obtain,

\begin{equation}
\label{eq1o} \langle I_{c,s} (x,t)\rangle^{(1)} =-i\sum_{\eta
\eta_1} \eta_1 \langle T_K\{I_{c,s} (x,t^\eta)\int dt_1
H^{bulk}(t_1^{\eta_1})\}\rangle.
\end{equation}

Starting from the expression (\ref{huu}) of the
Hamiltonian in terms of the bosonic fields and using the identity
$\lim_{\gamma
\rightarrow 0}(i\gamma)^{-1}\partial_t \text{exp}[i\sqrt{2}\gamma
\phi_c]=\sqrt{2}\partial_t \phi_c$, in order to cast the time
ordered averages into correlators of exponentials only,
we have:
\begin{eqnarray}
\label{1st}
&&\langle I_c (x,t)\rangle^{(1)}= -i\frac{e}{\pi}\sum_{n,m,n_s}
\frac{g^U_{n,m,n_s}}{(2\pi a)^n} \sum_{\epsilon=\pm} \sum_{\eta
\eta_1} \eta_1\int dt_1 \int_{-L/2}^{L/2} dx_1 e^{i\epsilon (\omega_0 t_1-\varphi_{n,m})}
e^{i\epsilon \Delta k_{n,m}x_1} \nonumber \\
&& \lim_{\gamma
\rightarrow 0}(i\gamma)^{-1}\partial_t \langle T_K \{e^{i\gamma
\sqrt{2} \phi_c(x,t^\eta)}e^{i \sqrt{2}\epsilon (n \phi_c
(x_1,t_1^{\eta_1})+n_s \phi_s (x_1,t_1^{\eta_1}))}\}\rangle=
\nonumber \\
&&=\frac{2e}{\pi} \sum_{n,m,n_s}
\left( \frac{L}{a}\right)^{-\frac{n^2K_c}{2}-\frac{n_s^2K_s}{2}}
\frac{g^U_{n,m,n_s} }{(2\pi a)^n}
\sum_{\eta \eta_1}\eta_1 \int dt_1 \int_{-L/2}^{L/2} dx_1 \sin ( \omega_0 t_1-\varphi_{n,m}
+\Delta k_{n,m}x_1) \partial_t G^{\phi_c \phi_c}_{\eta \eta_1}(x-x_1,t-t_1).\nonumber \\
\mbox{ }
\end{eqnarray}

\noindent where we have introduced $\epsilon=\pm$ for the
hermitian conjugates, and  the bosonic Keldysh Green's function is

\begin{equation}\label{green}
G^{\phi_c \phi_c}_{\eta \eta_1}(x-x_1,t-t_1)= <T_K \left( \sqrt{2}\phi_c
(x,t^\eta)\sqrt{2}\phi_c(x_1,t_1^{\eta_1})\right)> =
-\frac{K_c}{2} \sum_{\alpha=\pm}
\ln [a +i h_{\eta \eta_1} (t-t_1) ( v_c (t-t_1)  - \alpha (x-x_1))],
\end{equation}

\noindent with $\alpha=\pm$ for $R/L$ movers respectively, $h_{\pm \pm} (t)=\pm sgn(t), h_{\pm \mp} (t)=\mp 1$.
The non-trivial L dependence is arising from the correlator of the exponential
for a finite-size system.

Using the definition
of the Keldysh Green's function matrix elements and the symmetry property $G(x,\tau)=G(x,|\tau|)$,
only the terms with $\eta=-\eta_1$ can be retained, thus

\begin{equation}\label{curr1}
\langle I_c (x,t)\rangle^{(1)}= 2\frac{e}{\pi}
\sum_{n,m,n_s}\left( \frac{L}{a}\right)^{-\frac{n^2K_c}{2}-\frac{n_s^2K_s}{2}} n
\frac{g^U_{n,m,n_s}}{(2\pi a)^n} \sum_{\eta} \eta \int dt_1 \int_{-L/2}^{L/2} dx_1
\sin ( \omega_0 t_1 -\varphi_{n,m}+\Delta k_{n,m}x_1)
\partial_t G^{\phi_c \phi_c}_{\eta -\eta}(x-x_1,t-t_1).
\end{equation}
A further change of variables leads to final form of
first order contribution to the charge current,

\begin{equation}\label{ir}
\langle I_c (x,t)\rangle^{(1)}\propto
\sum_{n,m,n_s}\left( \frac{L}{a}\right)^{-\frac{n^2K_c}{2}-\frac{n_s^2K_s}{2}}
n\frac{g^U_{n,m,n_s}}{(2\pi a)^n} \sin (\omega_0 t-\varphi_{n,m}+ \Delta k_{n,m}x).
\end{equation}
and the spin current,
\begin{equation}\label{ir}
\langle I_s (x,t)\rangle^{(1)}\propto
\sum_{n,m,n_s}\left( \frac{L}{a}\right)^{-\frac{n^2K_c}{2}-\frac{n_s^2K_s}{2}}
n_s\frac{g^U_{n,m,n_s}}{(2\pi a)^n} \sin (\omega_0 t-\varphi_{n,m}+
\Delta k_{n,m}x).
\end{equation}

However, as these terms oscillate in space and in time no
pumping takes place to first order.  As we will show, to have a
d.c. current at least two umklapp operators with a non-zero phase
difference are required, in accordance with the general idea of pumping.

To second order we have:

\begin{equation}
\label{eq2o} \langle I_{c,s} (x,t)\rangle^{(2)} =-\frac 1 2
\sum_{\eta \eta_1\eta_2} \eta_1\eta_2 \langle T_K\{I_{c,s}
(x,t^\eta)\int dt_1 \int dt_2
H^{bulk}(t_1^{\eta_1})H^{bulk}(t_2^{\eta_2})\}\rangle.
\end{equation}

\noindent By using the bosonic expression of $H^{bulk}$ we find:

\begin{eqnarray}\label{2nd}
\langle I_c (x,t)\rangle^{(2)}=&& - \frac{e}{2\pi}
\sum_{n,m,n_s}\sum_{n',m',n_s'}
\frac{g^U_{n,m,n_s}}{(2\pi a)^n}\frac{g^U_{n',m',n_s'}}{(2\pi a)^{n'}}
\sum_{\epsilon_{1,2}=\pm}\sum_{\eta \eta_1 \eta_2} \eta_1 \eta_2
\int dt_1 \int dt_2 \int_{-L/2}^{L/2} dx_1 \int_{-L/2}^{L/2} dx_2  \nonumber \\
&& e^{i\epsilon_1 (\omega_0 t_1-\varphi_{n,m})}
e^{i\epsilon_2 (\omega_0 t_2-\varphi_{n',m'})}
e^{i\epsilon_1 \Delta k_{n,m}x_1} e^{i\epsilon_2 \Delta k_{n',m'}x_2} \times \nonumber \\
&& \lim_{\gamma \rightarrow 0} (i\gamma)^{-1} \partial_t \langle T_K
\left( e^{i\gamma \sqrt{2} \phi_c(x,t^\eta)}e^{i \epsilon_1\sqrt{2} (n \phi_c (x_1,t_1^{\eta_1}) +n_s \phi_s (x_1,t_1^{\eta_1}))}
e^{i \epsilon_2\sqrt{2} (n' \phi_c (x_2,t_2^{\eta_2})+{n'}_s\phi_s (x_2,t_2^{\eta_2}))}\right)\rangle.
\end{eqnarray}

A d.c. contribution to the current arises only from the term
with $\epsilon_1=-\epsilon_2$ and a non-zero phase difference. We
proceed to calculate it:

\begin{eqnarray}
\label{2nd1}
&&\langle I_c (x,t)\rangle^{(2)}_{d.c.}=\nonumber \\
&& =- \frac{e}{\pi}
\sum_{n,m,n_s}\sum_{n',m',n_s'}\left(\frac{L}{a}\right)^{-(n^2+n'^2)\frac{K_c}{2}-(n_s^2+n_s'^2)\frac{K_s}{2}}
\frac{g^U_{n,m,n_s}}{(2\pi a)^n}
\frac{g^U_{n',m',n_s'}}{(2\pi a)^{n'}} \sum_{\eta \eta_1 \eta_2}
\eta_1 \eta_2 \int dt_1 \int dt_2 \int_{-L/2}^{L/2} dx_1 \int_{-L/2}^{L/2} dx_2 \nonumber \\
&&\sin(\omega_0(t_1-t_2)-\Delta \varphi_{n,m}^{n',m'}+\Delta k_{n,m}x_1 -\Delta k_{n',m'}x_2)
 e^{-nn'G^{\phi_c \phi_c}_{\eta_1 \eta_2}(x_1-x_2,t_1-t_2)}
e^{-n_sn'_s G^{\phi_s \phi_s}_{\eta_1 \eta_2}(x_1-x_2,t_1-t_2)} \nonumber \\
&&\lbrack n \partial_t G^{\phi_c\phi_c}_{\eta \eta_1}(x-x_1,t-t_1) - n' \partial_t
G^{\phi_c \phi_c}_{\eta \eta_2}(x-x_2,t-t_2)\rbrack.
\end{eqnarray}

\noindent where $\Delta \varphi_{n,m}^{n',m'}=
(\varphi_{n,m}-\varphi_{n',m'})$ is the phase difference and the
Keldysh spin bosonic Green function is given by:

\begin{equation}\label{greens}
G^{\phi_s \phi_s}_{\eta \eta_1}(x-x_1,t-t_1)= <T_K \left( \sqrt{2} \phi_s
(x,t^\eta)\sqrt{2} \phi_s(x_1,t^{\eta_1})\right)>=-\frac{K_s}{2} \sum_{\alpha=\pm}
\ln [a +i h_{\eta \eta_1} (t-t_1) )  v_s (t-t_1)  - \alpha
(x-x_1))].
\end{equation}

An expression similar to (\ref{2nd1}) will hold for the spin current, except
that in this case the derivative of the spin bosonic Green's function
will appear, multiplied by $n_s$ (the spin umklapp quantum numbers),
instead of charge umklapp quantum numbers $n$.

The calculation of the contribution to the current from the
boundary terms is carried in an analogous way by considering
$H^{boundary}$ in Eqs.(\ref{eq1o})-(\ref{eq2o}) instead of
$H^{bulk}$.

\subsection{Zero temperature pumping}
\label{sec:ztp}

\subsubsection{bulk current}

Evaluating the integrals (\ref{1st})-(\ref{2nd}) (for details see
appendix), we find that the leading order contribution to the
charge and spin d.c. current at zero temperature is:

\begin{eqnarray}\label{cur}
I_c^{d.c.bulk}(\omega_0) &=& e K_cv_c
\sum_{n,m,n_s}\sum_{n',m',n'_s}(n-n') \left(
\frac{L}{a}\right)^{-(n-n')^2\frac{K_c}{2}-(n_s-n_s')^2\frac{K_s}{2}}
{\mathcal{A}}_{n,m,n_s}^{n',m',n_s'} I_{nmn_s}^{n'm'n_s'}
(\omega_0,\lbrack \Delta k_+ \rbrack_{ n,m}^{n',m'}) \frac{\sin
(\lbrack \Delta k_- \rbrack_{ n,m}^{n',m'})\frac{L}{2}}{\lbrack
\Delta k_- \rbrack_{ n,m}^{n',m'}}
\nonumber \\
I_s^{d.c.bulk}(\omega_0) &=& \hbar K_sv_s
\sum_{n,m,n_s}\sum_{n',m',n'_s}(n_s-n'_s)
\left(\frac{L}{a}\right)^{-(n-n')^2\frac{K_c}{2}-(n_s-n_s')^2\frac{K_s}{2}}
{\mathcal{A}}_{n,m,n_s}^{n',m',n_s'}
I_{nmn_s}^{n'm'n_s'}(\omega_0,\lbrack \Delta k_+ \rbrack_{
n,m}^{n',m'})
\frac{\sin (\lbrack \Delta k_- \rbrack_{ n,m}^{n',m'})\frac{L}{2}}{\lbrack \Delta k_-\rbrack_{ n,m}^{n',m'}}.\nonumber \\
\mbox{ }
\end{eqnarray}

\noindent where
$\lbrack \Delta k_\pm \rbrack_{ n,m}^{n',m'}=(\frac{\Delta k_{n,m}\pm \Delta k_{n',m'}}{2})$
and,
\begin{eqnarray}
{\mathcal{A}}_{n,m,n_s}^{n',m',n_s'}=\frac{g^U_{n,m,n_s}}{(2\pi a)^n}
\frac{g^U_{n',m',n'_s}}{(2\pi a)^{n'}} \sin \Delta \varphi_{n,m}^{n',m'}
\end{eqnarray}

\noindent is the area enclosed in a pumping cycle by the
periodic parameters $g^U_{n,m,n_s}(t)$ and $g^U_{n',m',n_s'}(t)$.
The expression  $I_{nmn_s}^{n'm'n_s'}$ for $v_c=v_s$ is given by,
(the case $v_s \ne v_c$ is treated in the appendix),

\begin{eqnarray}\label{ivcs}
I_{nmn_s}^{n'm'n_s'}(\omega_0,\lbrack \Delta k_+ \rbrack_{ n,m}^{n',m'})&=&
 \text{Sgn}(\omega_0)(\frac{a}{2v})^{2K^{n n'}_{n_s n'_s}}\Gamma^{-2}(K^{n n'}_{n_s n'_s})
\left(\omega_0^2-v^2\lbrack \Delta k_+^2 \rbrack_{ n,m}^{n',m'})\right)^{K^{n n'}_{n_s n'_s}-1}
\theta(|\omega_0|-|v\lbrack \Delta k_+ \rbrack_{ n,m}^{n',m'}|)  \nonumber \\
\mbox{ }
\end{eqnarray}

\noindent where $K^{n n'}_{n_s n'_s}=\frac{nn'K_c}{2}+\frac{n_sn'_sK_s}{2}$,
$K_c$ and $K_s$ are the
Luttinger parameters defined earlier, and the function
$\text{Sgn}(\omega_0) $ is defined as  $\text{Sgn}(\omega_0)=0 $ for $ \omega_0 =0 $ in addition to the usual definition
$\text{Sgn}(\omega_0) =\pm 1$ for $\omega_0$ positive/negative.

 The non trivial dependence of the current on the size of the pump $L$
arises technically from the exponential of the Keldysh correlators
evaluated on finite size of the pump with the usual ``charge
neutrality'' violated, $n \neq n' \;\;, n_s\neq n_s'$.  This violation is a
manifestation of the non equilibrium process taking place during the
pumping with ``charges'' in the upper part of the Keldysh contour not
canceling the charges in lower part.  Thus, the pumping can be viewed as
action of the potential on the section $L$ of the wire creating charge
unbalance and resulting in a net current in one
direction\cite{chamon_tunn}.

\subsubsection{discussion}

We now discuss the physical characteristics of our results. First,
the  nonlinear dependence on the size of the pumping region
strongly suppresses for large $L/a$ terms with large $|n-n'|$ or
$|n_s-n'_s|$.  Therefore, the leading contribution to the charge
current comes from terms with $n_s=n'_s$ and $n-n'=\pm1$, and the
leading contribution to the spin current comes from terms with
$n=n'$ and $n_s-n'_s=\pm 1$.  Second, depending on the lattice
having only charge umklapp terms (i.e. $n, n' \neq 0$ but $n_s
,n'_s =0$) or only spin umklapp terms ($n, n' =0$ but $n_s ,n'_s
\neq 0$), a pure charge or pure spin current will be induced. This
spin pumping takes place without spin-orbit coupling and without
magnetic field or spontaneous symmetry breaking, unlike the
mechanisms in Refs. [\onlinecite{fazio,chamon}]. This is possible
only due to interactions. Third, the charge and spin pumped per
cycle are not quantized but depend linearly on the area
${\mathcal{A}}_{n,m,n_s}^{n',m',n_s'}$ enclosed by the
interaction.  Note that at least two umklapp terms are needed to
have a non-zero d.c. current.  This accords with the observation
that at least two umklapp terms are required to represent a
lattice\cite{rosch}, and also in agreement with the picture that
electron pump is induced by the out of phase variation of any pair
of independent parameters. The current would vanish if under the
RG a single umklapp term is induced, even if associated with
several phases.  In case of mirror symmetry we have
$\varphi_{n,m},\varphi_{n',m'}=0$ resulting in a zero d.c.
current. Thus the breaking of mirror symmetry is a necessary
condition for the pumping.  Most importantly, the response of the
non Fermi-liquid (Luttinger) quasi particles to a fermionic
coupling produces anomalous frequency dependence in the pumped
current. Consider first the commensurate case where $\Delta
k_+=0$.  Eq.(22) reduces to a power law in frequency dependence
with an exponent $2(K^{n n'}_{n_s n'_s}-1)$.  In the
noninteracting limit, $K_c=1, K_s=1$, the lowest value of the
exponent will correspond to $K_{11}^{12} =3/2$, giving the
expected linear $\omega_0$ behavior at commensurability. In this
case we get charge and spin pumping with a frequency independent
pumping conductance,
${\mathcal{G}}_{c,s}=\frac{e^2}{h}\frac{2\pi}{\omega_0}I_{c,s}^{d.c.}$,
similar to Refs.\onlinecite{brouwer,chamon}.  With interaction,
the frequency dependence of the current is generally nonlinear
with an exponent depending on the strength of the Luttinger
interaction.  For $K^{n n'}_{n_s n'_s}>1$, the current goes to
zero smoothly in the zero frequency limit connecting to the
expected result of no current when the lattice does not oscillate.
In the range $K^{n n'}_{n_s n'_s}<1$, the Luttinger fixed point
would become unstable and a new CDW or SDW ground state forms,
where our considerations do not apply. This RG argument manifests
itself as a ``dynamic Stoner instability'' with $I(\omega_0)$
diverging as $\omega_0 \to 0$ in this case.  Note, however, the
stable regime includes both the superlinear and sublinear
behaviors in frequency dependences of the current.  Such
nontrivial power laws are never seen for conventional pumps.  In
the incommensurate case, the current vanishes in the frequency
window $|\omega_0|<|v\lbrack \Delta k_+ \rbrack_{ n,m}^{n',m'}|$.
This reflects the physical requirement that sufficient (photon)
energy must be supplied from the pumping source in order to make
the transition. The non-trivial power law appears again
immediately beyond the frequency threshold.

\subsubsection{boundary current}

We still need to examine the boundary contribution.  Carrying out the
 calculation along the lines descibed above we find that the mixed
 bulk-boundary contribution to the d.c. current vanishes while the
 pure boundary interference yields (cf. Ref. \onlinecite{chamon}),

\begin{equation}
\label{boundcur}
I_c^{d.c. boundary}=V_0^2\left(
\frac{L}{a}\right) ^{-K_{c}-K_{s}} |\omega
_{0}|^{K_{c}+K_{s}-1}\text{Sgn}(\omega _{0})\sin \varphi .
\end{equation}

We then conclude that for $L \to \infty$
(holding $\omega_0$ fixed so that no further renormalization of $V_0$
and $g^U_{nm}$  takes place) the bulk contribution
will dominate due to umklapps terms with $|n-n'| = 1 $ and $\omega_0 >
|v [\Delta k_+]^{n'}_n |$. The irrelevant terms acting over a large
distance win over the relevant terms from the edges.

\subsection{Finite temperature pumping}

Our consideration are easy to extend to small but finite temperature
(which leave the system in the vicinity of the Luttinger fixed
point). We start by considering the contribution from the bulk first.
Using the finite temperature expression for the correlation
functions of the boson operators\cite{schulz,giamarchi} the expression
for $I_{nmn_s}^{n'm'n_s'}(\omega_0, \lbrack \Delta k_+ \rbrack_{
n,m}^{n',m'})$ in (\ref{ivcs}) will read:

\begin{eqnarray}\label{curv}
&& I_{nmn_s}^{n'm'n_s',T}(\omega_0,\lbrack \Delta k_+ \rbrack_{ n,m}^{n',m'})= \nonumber \\
&&\left(\frac{2\pi a
T}{v}\right)^{2K^{n n'}_{n_s n'_s}-2}\sin(\pi K^{n n'}_{n_s n'_s})
 B(-\frac{i}{2\pi}s_++\frac{K^{n n'}_{n_s n'_s}}{2};1-K^{n n'}_{n_s n'_s})
B(-\frac{i}{2\pi}s_-+\frac{K^{n n'}_{n_s n'_s}}{2};1-K^{n n'}_{n_s n'_s})\sinh(\frac{\omega_0}{\pi T}),
\end{eqnarray}

\noindent where $s_\pm =\frac{(\omega_0\pm \lbrack v\Delta k_+ \rbrack_{ n,m}^{n',m'})}{2T}$
; $B(x,y)=\Gamma(x)\Gamma(y)/\Gamma(x+y)$ is
the Euler beta function.

When we consider incommensurate fillings, $\lbrack \Delta k_+
\rbrack_{ n,m}^{n',m'} \ne 0$, assuming  $T\ll v
\lbrack \Delta k_+ \rbrack_{ n,m}^{n',m'}$, two interesting regimes
occur depending on whether $T \ll \omega_0$, or $T \gg \omega_0$.
In the first case, we get:

\begin{eqnarray}\label{i}
&&I_{nmn_s}^{n'm'n_s',T}(\omega_0,\lbrack \Delta k_+ \rbrack_{ n,m}^{n',m'})
 \simeq \nonumber \\
&& \text{Sgn}(\omega_0) \sin(\pi K^{n n'}_{n_s n'_s})
\Gamma^2(1-K^{n n'}_{n_s n'_s})
\left(\frac{a}{2v}\right)^{2K^{n n'}_{n_s n'_s}-2}
\left(\omega_0^2-v^2\lbrack \Delta k_+^2 \rbrack_{ n,m}^{n',m'})\right)^{K^{n n'}_{n_s n'_s}-1}
\theta(|\omega_0|-|v\lbrack \Delta k_+ \rbrack_{ n,m}^{n',m'}|),
\end{eqnarray}
\noindent coinciding in the limit with the result at $T=0$. For $T
\gg \omega_0$ and  $\omega_0$ not too small compared to $v \lbrack
\Delta k_+ \rbrack_{ n,m}^{n',m'}$ we find:
\begin{eqnarray}\label{curt}
&&I_{nmn_s}^{n'm'n_s',T}(\omega_0,\lbrack \Delta k_+^2 \rbrack_{ n,m}^{n',m'})
\simeq \nonumber \\
&&\sin^2(\pi K^{n n'}_{n_s n'_s})\Gamma^2(1-K^{n n'}_{n_s n'_s})
\left(\frac{a}{2v}\right)^{2K^{n n'}_{n_s n'_s}-2}
\left(\omega_0^2-v^2\lbrack \Delta k_+^2 \rbrack_{ n,m}^{n',m'})\right)^{K^{n n'}_{n_s n'_s}-1}
e^{-\frac{v\lbrack \Delta k_+ \rbrack_{ n,m}^{n',m'}}{2T}}\sinh(\frac{\omega_0}{\pi T}),
\end{eqnarray}
\noindent where the exponential factor describes the suppression of processes between
initial and final states of energy $v|\lbrack \Delta k_+ \rbrack_{ n,m}^{n',m'}|/2$ involving
momentum transfer $\lbrack \Delta k_+ \rbrack_{ n,m}^{n',m'}$.
When $\omega_0 \to 0$ at low-temperature the exponential factor in (\ref{curt})
prevails and the processes with the smallest
$\lbrack \Delta k_+ \rbrack_{ n,m}^{n',m'}$ are favored and the current is suppressed.

At a typical commensurate point $\lbrack \Delta k_+ \rbrack_{ n_0,m_0}^{n_1,m_1}\sim 0$
and temperature not too low, we
have to balance algebraic and exponential suppression in (\ref{curt}).
In the limit $\omega_0\ll T$,
the dominant contribution to the d.c. current will be given by:

\begin{eqnarray}\label{wlt1}
&& I_c^{d.c.bulk}\sim \\ \nonumber && e K_cv_c n_0
{\mathcal{A}}_{n_0,m_0,n_{0s}}^{n_1,m_1,n_{1s}} \left(
\frac{\sin(\frac{G}{2n_0}L)}{\frac{G}{2n_0}}\right) \cos(\pi
K^{n_0 n_1}_{n_{0s} n_{1s}})^2 B^2(K^{n_0 n_1}_{n_{0s}
n_{1s}}/2,1-K^{n_0 n_1}_{n_{0s} n_{1s}}) \left(\frac{2 \pi a
T}{v}\right)^{2K^{n_0 n_1}_{n_{0s} n_{1s}}-2} \frac{\omega_0}{T}
\text{Sgn}(\omega_0).
\end{eqnarray}
In the non-interacting limit the lowest value of the exponent
corresponds to $K^{1 2}_{1 1}=3/2$ and one recovers again  the
usual Fermi liquid behavior $I \simeq max(T,\omega_0)$ for the
non-interacting gas\cite{giamarchi}.  With interactions present,
the current behaves as a power-law of the temperature with an
exponent depending on the interactions, indicating a strong
renormalization of the scattering process due to various
fluctuations of a one-dimensional electron gas. A similar
expression will hold also for the spin current with a coefficient
$K_sv_sn_{0s}$ instead of $K_cv_cn_0$.

\begin{figure}
\centerline{\includegraphics[angle=-0,width=\normwidth]{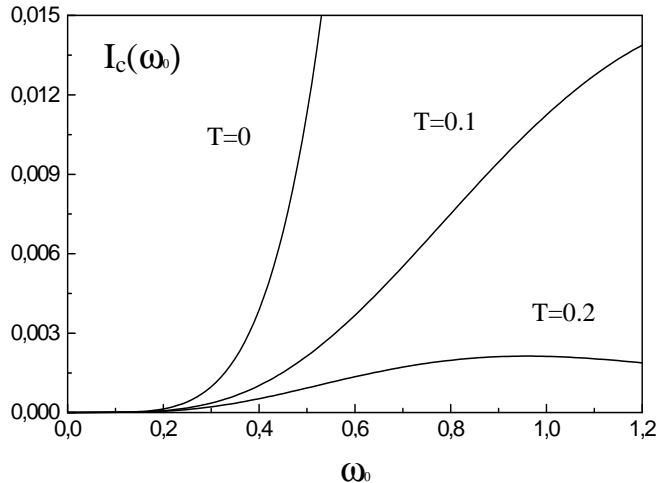}}
\vspace*{-8.cm}
\caption{The low-frequency behavior of the charge current
$I_c(\omega_0)$ at $T=0$ and $T=0.1,0.2$ having taken into account
umklapp terms $g_{2,0},g_{2,1},
g_{2,2},g_{3,1},g_{3,2},g_{3,3},g_{4,1},g_{4,2},g_{4,3}$.  We have
chosen $g_{n,m}=1$, $K_c=0.7$ and $K_s=1.$; $\omega_0$ and $T$ are
measured in units of $v \Delta k_{20}$ and $I(\omega_0)$ in units of
$ev_F/2$.}
\label{fig:totcurc}
\end{figure}

 Figure \ref{fig:totcurc} shows the low-frequency behavior of the charge current at zero and
finite temperature, taking into account few umklapp terms.

When considering the bulk-boundary contribution to the d.c.
current the same argument as in Sec.\ref{sec:ztp} holds. For $T\ll
\omega_0$ we recover the zero temperature expression
(\ref{boundcur})  and for $\omega_0\ll T$, we do have $I_c^{d.c.
boundary}\propto \left( \frac{L}{a}\right)^{-K_{c}-K_{s}}
|T|^{K_{c}+K_{s}-1}\text{Sgn}(\omega _{0})\sin \varphi$, so that
none of the previous conlcusions is invalidated when $L\rightarrow
\infty$ taking $\omega_0$ or $T$ fixed.

\section{Conclusions}\label{sec:conclusions}
We have introduced and studied charge and spin parametric pumps
for an interacting quantum wire.  We have demonstrated that the
pump, consisting of a  periodic potential oscillating in space and
in time over a size $L$ of a long clean wire, induces d.c. spin
and charge currents. At finite and fixed frequency, the leading
contribution to the  current arises from the interference of two
out-of-phase umklapp operators, in agreement with the picture of a
phase coherent quantum transport, while edges contribution
dominates at large but fixed size of the pump in the small
frequency limit. We have shown that the pumped current is strongly
affected by the interaction in the wire displaying a non-universal
behavior that depends on the filling and the interaction itself.
We have also discussed how to realize a pure spin pumping in the
wire as an alternative picture to the existing coherent spin
transport methods, without assuming any  magnetic field present.
We have finally addressed the question of the charge and spin
transported into a cycle across the section of the wire.  We have
shown that the charge and spin are not quantized even if the
adiabatic conditions are satisfied.

It would be interesting to address further questions regarding the
thermal current pumped into the system, dissipation and noise.
However, the most amusing question would concern the experimental
detection of our proposed pumping effect.

\acknowledgments We thank E. Andrei, E. Orignac, E. Shimshoni and in
particular A. Rosch for useful and enlightening discussions and to
L. Ioffe and L. Glazman for helpful comments on the manuscript.

\appendix

\section{Evaluation of the current integral to second order}

\subsection{Calculation at T=0}

In evaluating the current integral to second order (\ref{2nd}), due to time symmetry
properties of the Green's function, only the terms with $\eta_1=-\eta_2$, contribute. Changing variables we can
rewrite the integral (\ref{2nd1}) as:

\begin{eqnarray}\label{1st}
&&\langle I_c (x,t)\rangle^{(2)}= \nonumber \\
&& -\frac{ie}{\pi}
\sum_{n,m}\sum_{n',m'}\frac{g^U_{n,m}}{(2\pi a)^n}
\frac{g^U_{n',m'}}{(2\pi a)^{n'}} \sum_{\eta \eta_1}\int_{-L/2}^{L/2} dx_1 \int_{-L/2}^{L/2}
dx_2\int_{-\infty}^\infty dT
\sin( \omega_0 T-\Delta\varphi+ \Delta k_{n,m}x_1 -\Delta k_{n',m'}x_2)  \nonumber \\
&& e^{-nn'G^{\phi_c \phi_c}_{\eta_1 -\eta_1}(x_1-x_2,T)}
e^{-n_sn'_s G^{\phi_s \phi_s}_{\eta_1 -\eta_1}(x_1-x_2,T)}
\int_{-\infty}^\infty dT'  \lbrack n \partial_T
G^{\phi_c \phi_c}_{\eta \eta_1}(x-x_1,T')- n'
\partial_T G^{\phi_c \phi_c}_{\eta -\eta}(x-x_2,T')\rbrack,
\end{eqnarray}

\noindent where $\Delta \varphi=\Delta \varphi_{n,m}^{n',m'}$.

Another variable change $x_1,x_2\rightarrow
x=(x_1-x_2),x'=(x_1+x_2)$, so that we must evaluates the integrals
over $(x,T)$, $x'$ and $T'$, that we denote with $J_1,J_2,J_3$
respectively. The integral over $x'$ is simply
$J_2=2\int_{0}^{L/2} dx'\cos (\frac{\Delta k_{n,m} -\Delta
k_{n',m'}}{2}x')$. The integral over $(x,T)$ is:

\begin{equation}\label{int1}
J_1=\sum_{\eta_1} \int_{-\infty}^\infty dT\int_{-L/2}^{L/2} dx
\frac{ \sin \left( \omega_0 T-\Delta\varphi+ \frac{\Delta
k_{n,m}+\Delta k_{n',m'}}{2}x\right)}{\prod_{\alpha=\pm}\lbrack a
+ i h_{\eta_1 -\eta_1} (v_c T -\alpha x)\rbrack^{K_c^{nn'}}
\lbrack a + i h_{\eta_1 -\eta_1} (v_s T -\alpha
x)\rbrack^{K_s^{n_sn'_s}}}.
\end{equation}

\noindent where $K_c^{nn'}=nn'K_c/2$ and $K_s^{n_sn'_s}=n_sn'_sK_s/2$.

We first consider the case when $v_s=v_c$, and make the following
variable change $s=(vT-x)/v$, $s'=(vT+x)/v$, so that (\ref{int1}), becomes

\begin{equation}\label{int2}
\left(\frac{v}{2}\right)\sum_{\eta_1} \int_{-\infty}^\infty ds
\int_{-\infty}^\infty ds' \frac{\sin \left((\omega_0-v\Delta
k_+)s+(\omega_0+v\Delta k_+)s')+\Delta \varphi \right)}
{[a+2ih_{\eta_1 -\eta_1} vs]^{K^{n n'}_{n_s n'_s}}[a+2ih_{\eta_1
-\eta_1} vs']^{K^{n n'}_{n_s n'_s}}}
\end{equation}

\noindent where $K^{n n'}_{n_s n'_s}=K_c^{nn'}+K_s^{n_sn'_s}$,
$\frac{\Delta k_{n,m}+\Delta k_{n',m'}}{2}=\Delta k_+$,  indicated as
$\lbrack \Delta k_+ \rbrack_{ n,m}^{n',m'}$ in the main text.

Use the integrals 3.382.6/7 from reference \onlinecite{grad},
\begin{eqnarray}\label{intg1}
&&\int_{-\infty}^\infty  (\beta -i x)^{-\mu} e^{-ipx} dx= 2\pi \frac{e^{-\beta p}
(p)^{\mu-1}}{\Gamma (\mu)}\theta (p), \nonumber
\\ &&\int_{-\infty}^\infty (\beta +i x)^{-\mu}
e^{-ipx} dx= 2\pi \frac{e^{-\beta p} (-p)^{\mu-1}}{\Gamma
(\mu)}\theta (-p),
\end{eqnarray}
\noindent we find the final result of the main text (\ref{ivcs}).

When $v_s\ne v_c$ and $v_s < v_c$, we change variables\cite{gogolin_book},
$s=(v_sT+x)/(v_c+v_s)$ and $s'=(v_cT-x)/(v_c+v_s)$
 permitting us to rewrite the integral (\ref{int1}) as:

\begin{equation}
J_1=\sum_{\eta_1} \frac{(v_s-v_c)}{(v_s+v_c)} \int ds
\frac{e^{i\Delta \varphi}e^{-i\Omega s}} {\lbrack a + i h_{\eta_1
-\eta_1}(v_c+v_s)s\rbrack^{K_s^{nn'}}}F(s) -(s\rightarrow
-s;s'\rightarrow -s')
\end{equation}
\noindent where
\begin{eqnarray*}
F(s)=\int ds' \frac{e^{-i\Omega' s'}} {\lbrack a + i h_{\eta_1
-\eta_1}(v_c+v_s)s'\rbrack^{K_c^{nn'}} \lbrack a+i h_{\eta_1
-\eta_1} (2v_c s+(v_c-v_s)s')\rbrack^{K_c^{nn'}} \lbrack a+i
h_{\eta_1 -\eta_1} (2v_s
s'+(v_s-v_c)s)\rbrack^{K_s^{n_sn'_s}}},\nonumber
\end{eqnarray*}
\noindent and
\begin{eqnarray*}
\Omega &&=(\omega_0+v_c\Delta k_+) \\
\Omega'&&=(\omega_0-v_s\Delta k_+).
\end{eqnarray*}

\noindent We expect singularities in $I$ near $\omega_0=\pm v_s\Delta
k_+$ and $\omega_0=\pm v_c\Delta k_+$. Near
$\omega_0=v_s\Delta k_+$, $\Omega'\simeq 0$ and $\Omega=(v_c-v_s)\Delta k_+$.
The integral in $s$ is dominated by
$s<1/\Omega$ where $\Omega\simeq (v_c-v_s)\Delta k_+$, whereas
that in $s'$ is dominated by very large values. Power counting
does imply that

\begin{equation}\label{bo}
I(\omega_0) \sim \Theta (\omega_0-v_s\Delta k_+) (\omega_0
-v_s\Delta k_+)^{K_c^{nn'}+K_s^{n_s n'_s}-1}.
\end{equation}

\noindent Near $\omega_0=-v_c\Delta k_+$, the integrand in
$s'$ is dominated by $s'<1/\Omega'$, where $\Omega'\simeq
(v_c-v_s)\Delta k_+$ and that in $s$ by very large values. By
power counting we obtain the singular form of I:

\begin{equation}\label{bo1}
I(\omega_0) \sim \Theta (-\omega_0-v_c\Delta k_+) (-\omega_0
-v_c\Delta k_+)^{K_s^{n_s n'_s}+K_c^{nn'}-1}.
\end{equation}

The role of $v_c$ and $v_s$ will be exchanged if $v_c<v_s$.

In other ranges of $\omega_0$, the current may be written in terms of a
single integration
as shown in Ref.\onlinecite{meden}. We use
integrals 3.384.7/8 from Gradshteyn to perform first the integral
over $s$:
\begin{eqnarray}\label{intg}
&&\int_{-\infty}^\infty  (\beta -i x)^{-\mu}(\gamma -ix)^{-\nu}
e^{-ip} dx= 2\pi \frac{e^{-\beta p} (p)^{\mu+\nu-1}}{\Gamma
(\mu+\nu)}\Phi (\mu; \mu+\nu; (\beta-\gamma) p)  \theta (p),
\nonumber \\ &&\int_{-\infty}^\infty  (\beta +i x)^{-\mu}(\gamma
+ix)^{-\nu} e^{-ip} dx=- 2\pi \frac{e^{\beta p}
(-p)^{\mu+\nu-1}}{\Gamma (\mu+\nu)}\Phi (\mu; \mu+\nu;
(\beta-\gamma) p)  \theta (-p), \nonumber \\
\end{eqnarray}

\noindent  where $\Phi$ is the degenerate hypergeometric function,
and in our case $\beta=(a-ih_{\eta_1 -\eta_1}s')/(v_c-v_s)$ and
$\gamma=(a+i h_{\eta_1 -\eta_1}s')/(v_c+v_s)$.
Next, one employs the integral representation of the hypergeometric
function:

\begin{equation}
\Phi(a,b,z)=\frac{\Gamma(a)\Gamma(b)}{\Gamma(a+b)}
\int_0^1 ds e^{-zs}(1-z)^{b-a-1}s^{a-1}.
\end{equation}

\noindent The resulting integral over $s'$ may be written in terms of
the gamma functions by using $3.382.7$.
In the last step we use the integral 3.197.3 to recast the current in
terms of hypergeometric functions.

In the region $v_s \Delta k \le \omega_0 \le v_c \Delta k$ we obtain
the following result of the integral
$I_{nmn_s}^{n'm'n_s'}(\omega_0)$:
\begin{eqnarray*}
&& I_{n m n_s}^{n'm'n_s'}(\omega_0)=\nonumber \\
&& =\frac{2\pi a^K}{\Gamma(K)}
\frac{(v_c+v_s)^{1-K_s^{n_sn'_s}}}{(2v_c)^{K_c^{nn'}+1}}
(\omega_0-v_s\Delta k_+)^{K^{n n'}_{n_s n'_s}-1}(\omega_0+v_c \Delta k_+)^{K^{n n'}_{n_s n'_s}-1}
F\left(1,K_s^{n_sn'_s},K^{n n'}_{n_s n'_s};\frac{(v_c+v_s)}{2v_c}
\frac{\omega_0-v_s\Delta k_+}{\omega_0+v_c \Delta k_+}\right)\nonumber \\
\end{eqnarray*}
\noindent where $K^{n n'}_{n_s n'_s}=(K_c^{nn'}+K_s^{n_sn'_s})$.

\medskip
To have the current in its final form we must evaluate the
integral over $T'$ involving $\partial_{T'} G$.

\begin{equation}
\label{jc} J_3^c= \sum_{\eta_1} \int_{-\infty}^\infty dT' \lbrack
n
\partial_{T'} G^{\phi_c \phi_c}_{\eta \eta_1}(x-x_1,T')- n'
\partial_{T'} G^{\phi_c \phi_c}_{\eta -\eta_1}(x-x_2,T')\rbrack.
\end{equation}

\noindent for the charge current, where

\begin{eqnarray}
&& \sum_{\eta_1} \int_{-\infty}^\infty dT'\partial_{T'} G^{\phi_c
\phi_c}_{\eta \eta_1}(x-x_1,T') =-i\frac{K_cv_c}{2}
\sum_\alpha \sum_{\eta_1}
\int_{-\infty}^\infty dT' [ \frac{h_{\eta \eta_1}}{\lbrack
a+i h_{\eta \eta_1}\alpha (x-x_1) \rbrack +ih_{\eta
\eta_1}v_c T'}]=\nonumber \\
&&=\frac{K_cv_c}{2} \sum_\alpha\sum_{\eta_1}
\lim_{T_m \rightarrow \infty} \ln \lbrack a +i h_{\eta
\eta_1} (\alpha (x-x_1)+v_c T')\rbrack |_{-T_m}^{T_m} =
\nonumber \\ &&=\frac{K_cv_c}{2}\sum_\alpha \sum_{\eta_1} \lim_{T_m
\rightarrow \infty}\{\frac{1}{2} \ln \lbrack a^2 +h_{\eta
\eta_1}^2 (\alpha (x-x_1)+v_c T')^2\rbrack_{-T_m}^{T_m}+ i
\tan^{-1} ( h_{\eta \eta_1} \frac{(\alpha (x-x_1)+v_c T')}{a}
)|_{-T_m}^{T_m}\}=iK_cv_c\pi, \nonumber \\
\mbox{ }
\end{eqnarray}

\noindent and similarly for $ \int_{-\infty}^\infty dT'
\partial_{T} G^{\phi_c \phi_c}_{\eta -\eta_1}(x-x_2,T')$ with $\eta_1\rightarrow -\eta_1$.
\noindent We can perform the same type of calculation for the spin
bosonic Green's function. Thus we finally have:

\begin{equation}
\label{j1} J_3^c=iK_c v_c (n-n')\pi,
\end{equation}

\begin{equation}
\label{j2} J_3^s=iK_s v_s (n_s-n'_s)\pi.
\end{equation}

The result show that when $n_s',n_s=0$ we have a pure charge current $I_s=0$, while if
$n,n'=0$ we have a pure spin current. Since $J$ is a pure imaginary,
$I$ must be an immaginary too, to have a real quantity.

\subsection{Finite temperature}
For simplicity we make the calculation in the case $v_c=v_s$.
Using the finite temperature expression for the bosonic  Green's
function, we must evaluate the integral:
\begin{equation}\label{intt}
J_1=\sum_{\eta_1} \lbrack \frac{\pi a T}{v}\rbrack^{2K^{n n'}_{n_s
n'_s}} \int_{-\infty}^\infty dT\int dx \frac{\sin(\omega_0
T+\Delta k_+ x)} {\sinh\pi T \lbrack h_{\eta_1 -\eta_1} (\frac{v T
- x}{v})+ia \rbrack^{K^{n n'}_{n_s n'_s}} \sinh\pi T \lbrack
h_{\eta_1 -\eta_1} (\frac{v T + x}{v})-ia\rbrack^{K^{n n'}_{n_s
n'_s}}},
\end{equation}

\noindent  We first perform the variable change $s=vT-x$ and $s'=vT+x$ and
afterwards we use the integral:

\begin{equation}\label{sinh}
\int_{-\infty}^\infty ds |[\sinh (\pi T s)]|^{-K^{n n'}_{n_s n'_s}}e^{-isz}=
\frac{2^{K^{n n'}_{n_s n'_s}-2}}{\pi
T}B(\frac{K^{n n'}_{n_s n'_s}}{2}-\frac{iz}{2\pi},1-K^{n n'}_{n_s n'_s})
\cosh(\frac{z}{2T})[1+\tanh(\frac{z}{2T})],
\end{equation}

\noindent that permit us to write $I(\omega_0)$ in the final form shown in the
text.
The integral (\ref{jc}) gives the same result at finite temperature and similarly for
the spin part.

\end{document}